\documentclass[conference]{IEEEtran}
\usepackage[utf8]{inputenc}
\usepackage{cite}
\usepackage{graphicx}
\usepackage{amsmath}
\usepackage{mathrsfs,amsthm,amsmath,amssymb}
\usepackage{textcomp}
\usepackage{graphicx,multirow}
\usepackage{balance}
\usepackage{cite}
\usepackage{float}
\usepackage{gensymb}
\usepackage{eso-pic}
\usepackage{algorithm}
\usepackage{algorithmic}
\usepackage[hidelinks]{hyperref}
\usepackage{academicons}
\usepackage{xcolor}

\makeatletter
\newcommand\subparagraph{%
  \@startsection{subparagraph}{5}
  {\parindent}
  {3.25ex \@plus 1ex \@minus .2ex}
  {-1em}
  {\normalfont\normalsize\bfseries}}
\makeatother
\usepackage{titlesec}
\let\subparagraph\relax

\usepackage{titlesec}
\usepackage{scalerel}
\usepackage{tikz}
\usepackage{subcaption}

\usepackage[english]{babel}

\newcommand{\mb}{\mathbf}

\DeclareMathOperator{\tr}{Tr}

\usetikzlibrary{svg.path}

\definecolor{orcidlogocol}{HTML}{A6CE39}
\tikzset{
  orcidlogo/.pic={
    \fill[orcidlogocol] svg{M256,128c0,70.7-57.3,128-128,128C57.3,256,0,198.7,0,128C0,57.3,57.3,0,128,0C198.7,0,256,57.3,256,128z};
    \fill[white] svg{M86.3,186.2H70.9V79.1h15.4v48.4V186.2z}
                 svg{M108.9,79.1h41.6c39.6,0,57,28.3,57,53.6c0,27.5-21.5,53.6-56.8,53.6h-41.8V79.1z M124.3,172.4h24.5c34.9,0,42.9-26.5,42.9-39.7c0-21.5-13.7-39.7-43.7-39.7h-23.7V172.4z}
                 svg{M88.7,56.8c0,5.5-4.5,10.1-10.1,10.1c-5.6,0-10.1-4.6-10.1-10.1c0-5.6,4.5-10.1,10.1-10.1C84.2,46.7,88.7,51.3,88.7,56.8z};
  }
}

\newcommand\orcidicon[1]{\href{https://orcid.org/#1}{\mbox{\scalerel*{
\begin{tikzpicture}[yscale=-1,transform shape]
\pic{orcidlogo};
\end{tikzpicture}
}{|}}}}

\DeclareMathOperator{\st}{s.t.}
\DeclareMathOperator{\re}{Re}

\graphicspath{ {Images/} }
\IEEEoverridecommandlockouts

\begin{document}
\setlength{\parskip}{5pt}
\setlength{\abovedisplayskip}{5pt}
\setlength{\belowdisplayskip}{5pt}

% \title{Centralized and Distributed Learning in Satellite Communication Systems with Delayed CSI} 
\title{Tackling Delayed CSI in a Distributed Multi-Satellite MIMO Communication System}

% \title{Robust Distributed Space Massive MIMO Design with Delayed CSI} 
% \title{A Space-Air-Ground Integrated Network with Delayed CSI}

\author{Yasaman~Omid$^\text{\orcidicon{0000-0002-5739-8617}}$, 
Sangarapillai~Lambotharan$^\text{\orcidicon{0000-0001-5255-7036}}$,
Mahsa~Derakhshani$^\text{\orcidicon{0000-0001-6997-045X}}$
 
 \thanks{Yasaman Omid and Mahsa Derakhshani  are with the Wolfson School of Mechanical
Electrical and Manufacturing Engineering
 at Loughborough University, Loughborough, U.K. (e-mail: y.omid@lboro.ac.uk; M.Derakhshani@lboro.ac.uk).} 
 \thanks{Sangarapillai Lambotharan is with the Institute for Digital Technologies, Loughborough University London, London, UK (email: s.lambotharan@lboro.ac.uk) }
 % \thanks{Lojas Hanzo is with the School of Electronics and Computer Science, University of Southampton, Southampton, UK (email: lh@ecs.soton.ac.uk)}

}

\maketitle
\begin{abstract}

In this study, we explore the integration of satellites with ground-based communication networks. Specifically, we analyze downlink data transmission from a constellation of satellites to terrestrial users and address the issue of delayed channel state information (CSI). The satellites cooperate in data transmission within a cluster to create a unified, distributed massive multiple input, multiple output (MIMO) system. The CSI used for this process is inherently outdated, particularly due to the delay from the most distant satellite in the cluster. Therefore, in this paper, we develop a precoding strategy that leverages the long-term characteristics of CSI uncertainty to compensate for the undesirable impact of these unavoidable delays. Our proposed method is computationally efficient and particularly effective in lower frequency bands. As such, it holds significant promise for facilitating the integration of satellite and terrestrial communication, especially within frequency bands of up to $1$ GHz.

% In this paper, we study the integration of satellites with terrestrial communications. We consider a downlink transmission of multiple satellites towards users on the ground, under the consideration of delayed channel state information (CSI). The satellites in a cluster cooperate on signal transmission, to form a distributed massive MIMO communication system. The estimated CSI is delayed by the propagation delay of the furthest satellite in the cluster. By considering the availability of long-term properties of the CSI uncertainty, we design a precoding technique which can compensate for the undesirable effects of delayed CSI. The proposed technique has very low computational complexity and it works best for lower frequency bands. Therefore, using this technique can be helpful in integration of satellites with terrestrial communications using frequency bands of up to  $300$ MHz. 

\end{abstract}
\begin {IEEEkeywords}
LEO Satellite Communication, Delayed CSI, SAGIN
\end{IEEEkeywords}
\section{Introduction}\label{Section:Intro}
The integration of space, air, and ground networks (SAGIN) is a promising strategy for achieving universal connectivity, without the need to construct extra terrestrial base stations. Low Earth orbit (LEO) satellites are important in this setup, providing coverage for remote and under-served regions. These satellites, stationed between $500$km and $2000$km above the Earth, move swiftly, requiring quick and efficient handovers to maintain constant service for users. In mega-constellations like Starlink, multiple satellites are visible to a user terminal on the ground. In other words, a given user can be connected to multiple satellites which are within its $30^{\degree}$ elevation angle.  It has been shown in the literature that to substantially enhance the system performance in mega-constellations, the satellites could collaborate on signal detection in the uplink \cite{OmidSpaceMIMO,10293747} as well as precoding and signal transmission in the downlink \cite{9939157,9265189}. In addition to the performance enhancements, multi-satellite collaboration offers the added advantage of redundancy; i.e. in case the link to one satellite becomes obstructed, other satellites can step in to compensate for the loss.

This paper studies a network where a constellation of LEO satellites provides service for users in isolated zones without ground infrastructure. Users in these areas have a direct line of sight (LOS) links to multiple satellites at any time. These satellites are connected to each other by their inter-satellite links. In the uplink, we use time-division duplexing (TDD) so that users could send distinct pilot sequences for satellites to estimate the channel state information (CSI). This information is then processed to shape an effective downlink precoding matrix.
One of the major problems in satellite communications (SatCom) networks is the delay caused by the large distance between satellites and ground users. This delay, extending beyond the coherence intervals of channels, means that the CSI obtained during uplink does not match the actual downlink conditions. 
This issue, referred to as the ``delayed CSI problem"  \cite{9439942}, is a key challenge in downlink communications of SatCom networks, irrespective of whether TDD or frequency-division duplexing (FDD) is used. Our study concentrates on TDD due to its lower overhead benefits.

In the literature, many works have focused on the delayed CSI problem in downlink Satcom. 
In \cite{9439942}, the delayed CSI problem was investigated for a massive multiple input multiple output (MIMO) satellite communication, where one satellite was equipped with a large antenna array. The authors studied the massive MIMO channel model by including the characteristics of LEO satellites, and then they proposed a deep learning (DL)-based channel predictor which eliminated the impact of the delayed CSI.
The authors of \cite{9826890} adopt a hybrid beamforming technique for data transmission in a satellite equipped with large antenna arrays. In hybrid beamforming, the number of radio frequency (RF) chains is much smaller than the number of antennas, which results in reduced hardware complexity. However, in this case, the number of users served by the satellite must not exceed the number of the RF chains. Two deep neural network (DNN) systems are used, one for CSI prediction and the other for hybrid beamforming. In the first DNN, the downlink channels are predicted using the correlations of the uplink and the downlink CSI in both TDD and FDD modes. Then, using the second DNN the predicted channels are mapped to a desirable hybrid beamformer.
Further research on channel estimation for massive MIMO LEO satellite systems is given in \cite{10200015,10208031,10008701,10008605}.

While many papers have investigated the effects of delayed CSI in massive MIMO LEO SatCom, to the best of our knowledge this problem is not yet investigated in distributed massive MIMO satellite communications. In addition, most of the works in this area consider use of DL-based techniques, which may not be applicable to satellites that are already in-orbit and do not possess the necessary infrastructure. Motivated by the above, in this paper we propose a low-complexity, robust precoding technique for distributed MIMO SatCom to minimize the harmful effects of delayed CSI. Thus, we optimize the precoding matrix for all satellites through solving an mean squared error (MSE) minimization problem. We assume some of the long-term properties of the CSI uncertainty are available to the satellites and the delayed instantaneous CSI of all satellites is shared in the cluster. By obtaining these information, one of the satellites with better signal processing capabilities acts as a network controller (NC) and calculates the optimal precoding matrix. This satellite then needs to share the precoding information to all satellites in the cluster. Our results show that while this method is useful and computationally efficient, it can compensate for the effects of delayed CSI up to a certain frequency. By increasing the frequency, the changes in channels becomes more variable with time; hence, more complex channel prediction techniques are needed. However, the proposed algorithm is useful for frequency bands of up to {$1$ GHz}, meaning that it can integrate with many existing terrestrial communication systems.

The rest of this paper is organized as follows. Section \ref{Section: System Model} is dedicated to system model and channel model. In Section \ref{Section: Problem Formulation} the problem is formulated and in Section \ref{Section Proposed Solution} the solution is presented. Our simulation setup and results are demonstrated in Section \ref{Section Numerical Results}, and finally Section \ref{Section Conclution} concludes the paper.

\begin{figure}[t]
    \centering
    \includegraphics[scale=0.5]{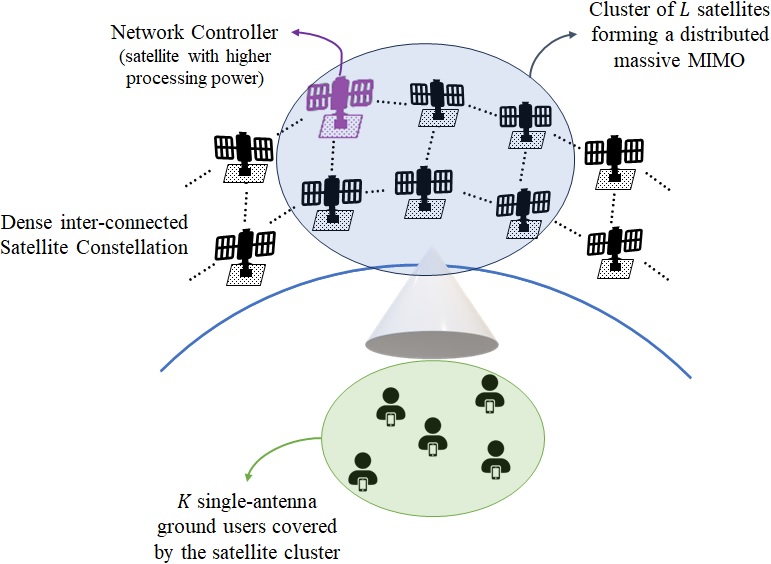}
    \caption{Distributed space massive MIMO system model.}
    \label{fig:system model}
\end{figure}

\section{System Model}\label{Section: System Model}
We examine a densely populated LEO satellite constellation, assuming inter-satellite links (ISLs) for connectivity between satellite nodes. In this context, we assume that ISLs form high-capacity communication channels with no inter-satellite interference. The system model, depicted in Fig. \ref{fig:system model} and inspired by the work in \cite{9939157}, presents this setup.
Each cluster comprises $L$ satellites or airborne vehicle nodes, forming a distributed massive MIMO network. Satellite mobility may cause some satellites to go out of sight and this necessitates their replacement by new satellites. This network of satellites establishes connectivity for a designated off-the-grid terrestrial region, accommodating $K$ mobile users equipped with single antennas.

Through TDD, the channel estimation is conducted in the uplink and CSI estimates are then utilized to design a precoding technique for downlink transmission. Given the substantial distances between satellites and users, coupled with limited coherence intervals, we confront the challenge of delayed CSI. Our paper addresses this issue within the framework of a centralized processing technique.
In this configuration, we designate a satellite within the cluster with considerable processing capabilities, to function as the NC. Centralized processing becomes advantageous when leveraging an existing constellation of older satellites, each with limited processing capabilities, to jointly cover a specific area. This way, only a small fraction of satellites in the constellation needs to be equipped with complex infrastructures.

% \textcolor{blue}{(Channel model, including CSI uncertainty model should be brought here:)}

Here we assume that each satellite in the cluster designates $M$ antennas for covering this specific area on the ground. Thus, the cluster of satellites can be thought of as a distributed access point with $ML$ antennas.
The channels between users and satellites are defined as $\mb{G}\in\mathbb{C}^{ML\times K}$ with each element following a Rician distribution. The channels are considered to be time-variant and they are based on the works in \cite{9439942,9110855,9849035,9628071}. Particularly, the channel vector from the $l$th satellite to the $k$th user, i.e. $\mb{g}_l^{[k]}\in\mathbb{C}^{M\times 1}$, can be modelled by 
\begin{equation} \label{Channel Model}
    \mb{g}_l^{[k]}=\frac{1}{\text{PL}_l^{[k]}}\left(\mb{h}_{\text{LOS},l}^{[k]}+\mb{h}_{\text{NLOS},l}^{[k]}\right),
\end{equation}
where $\text{PL}_l^{[k]} = \frac{4 \pi d_l^{[k]}f}{c}$ is the path-loss, with $d_l^{[k]}$ as the distance between the $l$th satellite and the $k$th user, $f$ as the frequency and $c$ as the speed of light. The terms $\mb{h}_{\text{LOS},l}^{[k]}$ and $\mb{h}_{\text{NLOS},l}^{[k]}$ stand for the LOS and non-line of sight (NLOS) properties of the channel defined by (\ref{h_LOS}) and (\ref{h_NLOS}) on top of the next page. The definitions of the parameters in (\ref{h_LOS}) and (\ref{h_NLOS}) are given in Table \ref{tab: params}.
\begin{table}[t]
    \centering
    \begin{tabular}{|c|c|c|}
    \hline
        \textbf{Parameter} & \textbf{Definition}  \\
        \hline
         $\kappa^{[k]}_{l}$& Rician factor, uniform random variable $\in(80,90]$  \\ \hline
         $P^{[k]}$ & Num. NLOS paths, uniform random variable $\in(1,7]$ \\ \hline
         $\bar{v}_{\text{Sat},l}^{[k]}$ & LOS satellite Doppler frequency, $\bar{v}_{\text{Sat},l}^{[k]}=\frac{q_l}{c}f\cos(\omega_l^{[k]})$ \\ \hline
         $q_l$ & Satellite velocity, $q_{l}=\sqrt{\frac{GM_e}{R+\text{Altitude}_{l}}}$, \\ & $G=6.674\times 10^{-11}$($\frac{\text{m}^3}{\text{kg}\text{s}^{2}}$) is gravitational constant, \\ & $M_e=5.972\times 10^{24}$(kg) is mass of the Earth, \\ & $R=6.371\times 10^6$(m) is radius of Earth \\ & $\text{Altitude}_{l}$ is the satellite altitude \\ \hline
         $\omega_l^{[k]}(t)$ & Angle between the  satellite's forward velocity and \\ & the boresight from satellite to user \\ \hline
         $\bar{v}_{\text{UE},l}^{[k]}$ & LOS user Doppler frequency \\ \hline
         $v_{\text{Sat},l,p}^{[k]}$ & NLOS satellite Doppler frequency, \\
          & $v_{\text{Sat},l,p}^{[k]} = \bar{v}_{\text{Sat},l}^{[k]}$, $\forall p\in\{1,...,P^{[k]}\}$\\ \hline
         $v_{\text{UE},l,p}^{[k]}$ & NLOS user Doppler frequency for path $p$\\ \hline
         $\tau_{\text{LOS},l}^{[k]}$ & Delay of LOS path, $\tau_{\text{LOS},l}^{[k]} =\frac{ d_l^{[k]}}{c}$  \\ \hline
         $\theta_{\text{LOS},l}^{[k]}$ & Angle of horizontal direction for the LOS path  \\ \hline
         $\psi_{\text{LOS},l}^{[k]}$ & Angle of vertical direction for the LOS path \\ \hline
         $g^{[k]}_p$ & Gain of the $p$th NLOS path, $g^{[k]}_p\sim\mathcal{CN}(0,1)$ \\ \hline
         $\tau_{\text{NLOS},l,p}^{[k]}$ & Delay of the $p$th NLOS path, $\tau_{\text{NLOS},l,p}^{[k]}=\frac{d_{l,p}^{[k]}}{c}$ \\ \hline
         $\theta_{\text{NLOS},l,p}^{[k]}$ & Angle of horizontal direction for the $p$th NLOS path,\\ & Randomly generated\\ \hline
         $\psi_{\text{NLOS},l,p}^{[k]}$ & Angle of vertical direction for the $p$th NLOS path,\\ & Randomly generated \\ \hline
         $\mb{u}(\theta,\psi)$ & Array response vector, \\ & $\mb{u}(\theta,\psi)=\mb{a}\left(\cos(\theta)\sin(\psi),M_x\right)\otimes \mb{a}\left(\cos(\psi),M_y \right)$ \\ \hline 
         $\mb{a}(\phi,N)$ & One-dimensional steering vector function, \\  & $\mb{a}(\phi,N)=\frac{1}{\sqrt{N}}[1,e^{-j2\pi\frac{d_s}{\lambda} \phi},...,e^{-j2\pi\frac{d_s}{\lambda}(N-1) \phi}]^T$ \\ \hline
         $\lambda$ & Carrier wavelength $\lambda=\frac{c}{f}$ \\ \hline
         $d_s$ & Antenna spacing (considered to be half a wavelength) \\ \hline
    \end{tabular}
    \caption{Parameter definitions of the channel between the $l$th satellite and the $k$th user.}
    \label{tab: params}
\end{table}

\begin{figure*}
    \begin{align}\label{h_LOS}
    &\mb{h}_{\text{LOS},l}^{[k]}=\sqrt{\frac{\kappa^{[k]}_{l}}{1+\kappa^{[k]}_{l}}}\times exp\left(j2\pi t\left(\bar{v}_{\text{Sat},l}^{[k]}+\bar{v}_{\text{UE},l}^{[k]}\right)\right)\times exp\left(-j2\pi f\tau_{\text{LOS},l}^{[k]} \right)\times\mb{u}\left(\theta_{\text{LOS},l}^{[k]},\psi_{\text{LOS},l}^{[k]} \right),\\
\label{h_NLOS}
    &\mb{h}_{\text{NLOS},l}^{[k]}=\sqrt{\frac{1}{P^{[k]}\left(1+\kappa^{[k]}_{l}\right)}}\times\sum_{p=1}^{P^{[k]}}g^{[k]}_p\times exp\left(j2\pi t\left(v_{\text{Sat},l,p}^{[k]}+v_{\text{UE},l,p}^{[k]}\right)\right)
    \nonumber\\&\qquad\qquad\qquad\qquad\qquad\qquad\quad\quad\quad\qquad
    \times exp\left(-j2\pi f\tau_{\text{NLOS},l,p}^{[k]} \right)\times\mb{u}\left(\theta_{\text{NLOS},l,p}^{[k]},\psi_{\text{NLOS},l,p}^{[k]} \right),
\end{align}
\end{figure*} 
% Specifically, the statistical properties of the channel based on the Rician distribution are presented in \cite{9628071}.

% \textcolor{blue}{(we can bring the Rician distribution of the channel here instead of just citing the reference.)}

Due to the channel variations through time, and the delayed CSI estimation, the estimated channel gain $\mb{\hat{G}}$ can be written as 
\begin{equation}
    \mb{\hat{G}}=\mb{G}+\mb{\Tilde{G}},
\end{equation}
where $\mb{\Tilde{G}}$ represents the estimation error caused by the delay, and its value is affected by the operating frequency and satellite velocity.

The value of $\mb{\Tilde{G}}$ highly depends on the operating frequency, maximum distance of satellites in the cluster from a user, and the satellites' maximum velocity. In a real scenario, at a given time $t$, the estimated CSI  $\mb{\hat{G}}(t)$ is in fact the CSI of $T_d$ seconds ago, e.g. {$\mb{\hat{G}}(t)=\mb{G}(t-T_d)$}. The estimation delay is calculated by $T_d=\frac{d_{\text{max}}}{c}$, where $d_{\text{max}}$ is the maximum distance between satellites in the cluster and users on Earth. On the other hand, by increasing the operating frequency, the satellite Doppler frequency also increases, i.e. $\bar{v}_{\text{Sat},l}^{[k]}=\frac{q_l}{c}f\cos(\omega_l^{[k]})$. Based on (\ref{h_LOS}) and (\ref{h_NLOS}), we can see that the variation of the exponential terms w.r.t time depends highly on the satellites' Doppler frequency. In other words, a small change in time would change the term $e^{j2\pi t \bar{v}_{\text{Sat},l}^{[k]}}$ more if $\bar{v}_{\text{Sat},l}^{[k]}$ is larger. Thus, the value of $\mb{\Tilde{G}}(t)=\mb{G}(t-T_d)-\mb{G}(t)$ becomes larger by increasing the frequency. In our work, we compensate for a tolerable CSI error (for frequency ranges of up to $1$ GHz) by using signal processing techniques. In higher frequencies, on the other hand, other techniques should be considered, such as deep learning tools to leverage the channel correlations through time for more accurate channel predictions based on the estimated delayed CSI \cite{9439942}.

\section{Problem Formulation}\label{Section: Problem Formulation}
We consider the downlink transmission of symbols $\mb{s}=[s^{[1]},...,s^{[K]}]^T$ from the cluster of satellites to $K$ users on the ground. Using a precoding matrix $\mb{V}\in\mathbb{C}^{ML\times K}$, the symbols are precoded  as $\mb{x}=\mb{V}\mb{s}$ before transmission. Then, the signal $\mb{x}\in\mathbb{C}^{ML\times 1}$ is transmitted by all satellites in the cluster simultaneously. Note that each satellite only transmits a $M\times 1$ part of the signal $\mb{x}$. The received signal by a given user $k$ is demonstrated by $y^{[k]}$ which is the $k$th element of the vector $\mb{y}$ 
\begin{equation}
    \mb{y} = \mb{G}^H\mb{x}+\mb{n} = \mb{G}^H\mb{V}\mb{s}+\mb{n},
\end{equation}
where $\mb{n}$ is the additive white Gaussian noise following the distribution $\mb{n}\sim\mathcal{CN}(\mb{0},\sigma^2\mb{I})$.
To cope with interference and the destructive effects of the delayed CSI, we study the precoding matrix optimization aiming to minimize the MSE of the received signals. Without loss of generality, assuming $\mathbb{E}\{\mb{s}\mb{s}^H\}=\mb{I}$, the MSE is calculated by
\begin{align}
    \text{MSE} &= \mathbb{E}\{(\mb{y}-\mb{s})^H(\mb{y}-\mb{s})\}
    \nonumber\\&=\tr(\mb{V}^H\mathbb{E}\{\mb{G}\mb{G}^H\}\mb{V})-2\re\{\tr(\mb{V}^H\mathbb{E}\{\mb{G}\})\}\nonumber\\&\quad+K(\sigma^2+1).
\end{align}
Now, by replacing $\mb{G}=\mb{\hat{G}}-\mb{\Tilde{G}}$, we have 
\begin{align}
    \text{MSE} =& \tr(\mb{V}^H\mb{\hat{G}}\mb{\hat{G}}^H\mb{V})+\tr(\mb{V}^H\mathbb{E}\{\mb{\tilde{G}}\mb{\tilde{G}}^H\}\mb{V})\nonumber\\&
    -2\re\{\tr(\mb{V}^H\mathbb{E}\{\mb{\tilde{G}}\}\mb{\hat{G}}^H\mb{V})\}-2\re\{\tr(\mb{V}^H\mb{\hat{G}})\}\nonumber\\&
    -2\re\{\tr(\mb{V}^H\mathbb{E}\{\mb{\tilde{G}}\})\}+K(\sigma^2+1).
\end{align}
We formulate the optimization problem based on a MSE minimization technique, as
\begin{equation}\label{main opt}
\begin{array}{cl}
\min\limits_{\mb{V}}& \displaystyle\text{MSE}\\ 
\st 
&\tr \left(\mb{V}_l\mb{V}_l^H\right)\leq P_{l}, \ l\in\{1,...,L\}, \\
\end{array}
\end{equation}
where $P_{l}$ is the power budget of the $l$th satellite, and $\mb{V}_l\in\mathbb{C}^{M\times K}$ is a sub-precoding matrix for the $l$th satellite. Particularly, $\mb{V}_l$ is a matrix containing the rows $\left((l-1) M+1\right)$ to $(M l)$ of the precoding matrix $\mb{V}$.

By solving this optimization problem, the precoding matrix $\mb{V}$ can be calculated. This solution is entirely given by the NC satellite. The NC needs to access the following information about other satellites in the cluster: i) the instantaneous estimated CSI of all satellites, and ii) the distribution of CSI uncertainty for all satellites. In turn, the NC shares the precoding sub-matrices $\mb{V}_l$, $l\in\{1,...,L,\ l\neq \text{NC}\}$, with all other satellites. In this context, we assume that the satellites in a cluster are perfectly synchronized and they are capable of simultaneous data transmission. Note that as we are considering a dense constellation, the largest delay in the cluster does not have much difference with the smallest delay which belongs to the satellite at $90^{\degree}$ elevation angle; thus, synchronization is more efficiently performed in the sense that  the idle time of the nearer satellites is small. In the next section we propose a solution for this optimization problem.

\section{Proposed Solution}\label{Section Proposed Solution}
A key factor in developing a practical solution for the optimization problem in (\ref{main opt}) is maintaining the computational complexity low. Complex solutions take longer and the longer the processing takes, the more the channel will change. Since the constraint of this optimization problem is based on each individual satellite, the problem in (\ref{main opt}) can be decomposed into $L$ sub-problems. Through the following considerations of $\mb{V}=[\mb{V}_1^T,...,\mb{V}_L^T]^T$, $\mb{\hat{G}}=[\mb{\hat{G}}_1^T,...,\mb{\hat{G}}_L^T]^T$ and $\mb{\tilde{G}}=[\mb{\tilde{G}}_1^T,...,\mb{\tilde{G}}_L^T]^T$,
the sub-problem for the $l$th satellite is given by 
\begin{equation}\label{sub-problem}
\begin{array}{cl}
\min\limits_{\mb{V}_l}& \displaystyle f(\mb{V}_l) \\ 
\st 
&\tr \left(\mb{V}_l\mb{V}_l^H\right)\leq P_{l}, \\
\end{array}
\end{equation}
The objective function of (\ref{sub-problem}) is
\begin{align}
    f(\mb{V}_l) =\tr\left(\mb{V}_l^H\mb{A}_l\right)-2\re\left\{\tr\left(\mb{V}_l^H(\mb{\hat{G}}_l+\mathbb{E}\{\mb{\Tilde{G}}_l\})\right)\right\},
\end{align}
where
\begin{align}
    \mb{A}_l = \sum_{i=1}^L& \mb{\hat{G}}_l \mb{\hat{G}}_i^H \mb{V}_i + \mathbb{E}\{\mb{\Tilde{G}}_l \mb{\Tilde{G}}_i^H\} \mb{V}_i  \nonumber\\&
    - \mathbb{E}\{\mb{\Tilde{G}}_l\} \mb{\hat{G}}_i^H \mb{V}_i -\mb{\hat{G}}_l \mathbb{E}\{\mb{\Tilde{G}}_i^H\} \mb{V}_i.
\end{align}

The solution to this optimization problem can be given by an alternating algorithm where in each step the problem is solved for only one satellite and all other sub-precoding matrices of other satellites are considered constant. The constraint is enforced using the Lagrangian method, i.e. the term $\lambda_l\left(\tr(\mb{V}_l\mb{V}_l^H)-P_l\right)$ is added to the objective function of each sub-problem, where $\lambda_l$ is the dual Lagrangian variable. This results in the following unconstrained optimization problem 
\begin{equation}\label{sub-problem_lagrangian}
\begin{array}{cl}
\min\limits_{\mb{V}_l}& \displaystyle f(\mb{V}_l)+\lambda_l\left(\tr(\mb{V}_l\mb{V}_l^H)-P_l    \right).
\end{array}
\end{equation}
In each step of this algorithm, at first the optimal values for $\mb{V}_l$, $l\in\{1,...,L\}$ are obtained using the first-order optimality condition.
The optimal solution, in this case, is given by (\ref{V_l solution}) on top of the next page.
\begin{figure*}[t]
\begin{align}\label{V_l solution}
    \mb{V}_l = &\Bigg( \mb{\hat{G}}_l \mb{\hat{G}}_l^H + \mathbb{E}\{\mb{\Tilde{G}}_l \mb{\Tilde{G}}_l^H\} - \mathbb{E}\{\mb{\Tilde{G}}_l\} \mb{\hat{G}}_l^H -\mb{\hat{G}}_l \mathbb{E}\{\mb{\Tilde{G}}_l^H\}  +\lambda_l\mb{I}\Bigg)^{-1}
    \nonumber\\&
    \quad\times \Bigg( \mb{\hat{G}}_l- \mathbb{E}\{\mb{\Tilde{G}}_l\}  - \sum_{i\neq l}^L \left( \mb{\hat{G}}_l \mb{\hat{G}}_i^H \mb{V}_i + \mathbb{E}\{\mb{\Tilde{G}}_l \mb{\Tilde{G}}_i^H\} \mb{V}_i \right)   +\sum_{i\neq l}^L \left( \mathbb{E}\{\mb{\Tilde{G}}_l\} \mb{\hat{G}}_i^H \mb{V}_i +\mb{\hat{G}}_l \mathbb{E}\{\mb{\Tilde{G}}_i^H\} \mb{V}_i \right) \Bigg).
\end{align}
\end{figure*}
After updating the sub-precoding matrix for each satellite, the dual Lagrangian variable for each satellite, e.g. $\lambda_l$, $l\in\{1,...,L\}$, should be updated. 
Based on the above, the algorithm for solving (\ref{sub-problem_lagrangian}) in an iterative manner is given in Algorithm \ref{Alg11}. Note that, all these computations take place in the NC satellite, and during the iterations there is no information sharing required within the cluster. The Algorithm continues until a sum-rate-based convergence criteria is met.

\begin{algorithm} [t]
\caption{Alternating optimization algorithm for solving the problem in (\ref{sub-problem_lagrangian}).}
\begin{algorithmic} \label{Alg11}
\STATE $\mb{Initialize }$  \ $\mb{V}$ such that the constraints are met
\STATE $\mb{Initialize }$ $\lambda_l = 1$ $\forall l\in\{1,...,L\}$
\STATE $\mb{Initialize }$ $\alpha = 0.9$ and $i=0$
\STATE $\mb{Repeat}$
\STATE \quad $i=i+1$
\STATE \quad \textbf{for} $l\in\{1,...,L\}$

\STATE \quad \quad 1. Compute $\mb{V}_l$  by (\ref{V_l solution})
\STATE \quad \quad 2. if $\tr(\mb{V}_l\mb{V}_l^H)>P_l$\\
\STATE \quad \quad\quad \quad $\lambda_l = \lambda_l + \alpha (\tr(\mb{V}_l\mb{V}_l^H)-P_l)$\\
\STATE \quad\quad\quad else
\STATE \quad \quad\quad \quad $\lambda_l = \alpha \lambda_l$
\STATE $\mb{Until}$ $|\text{Rate}(i)-\text{Rate}(i-1)|<0.01\times\text{Rate}(i)$
\end{algorithmic}
\end{algorithm}

In Algorithm \ref{Alg11}, the termination point is decided by the convergence of the sum rate, which is calculated by
\begin{equation}
\text{Rate} =\sum_{k=1}^{K}\log\left(1+\frac{|{\mb{g}}^{[k]H}\mb{v}^{[k]}|^2}{\sum_{i\neq k}^{K}|{\mb{g}}^{[k]H}\mb{v}^{[i]}|^2+\sigma^2} \right),
\end{equation}
where ${\mb{g}}^{[k]}\in\mathbb{C}^{ML\times 1}$ is the $k$th column of the channel matrix $\mb{G}$ dedicated for the channels from all satellites in the cluster to the $k$th user, and similarly $\mb{v}^{[k]}\in\mathbb{C}^{ML\times 1}$ is the $k$th column of the precoding matrix $\mb{V}$.

Based on Algorithm 1, the computational complexity of the algorithm in each iteration contains $L$ computations of $\mb{V}_l$, $L$ updates for the dual Lagrangian variable, and one sum-rate calculation. The majority of the complexity lies in calculating $\mb{V}_l$. Thus, the computational complexity of each iteration of Algorithm \ref{Alg11} is in the order of $\mathcal{O}\left(L\left(13M^2K+4M^2+3(L-1)MK\right) \right)$. Note that, the order of complexity grows polynomially with $L$ and $M$, and linearly with $K$. Furthermore,
based on our simulations, the algorithm converges on around 10 iterations. Thus, the total complexity of this algorithm, while being iterative, is quite low. 

%\textcolor{orange}{\\  talk about the convergence of the algorithm }

\begin{figure*}[t]
    \centering
    \includegraphics[scale=0.28]{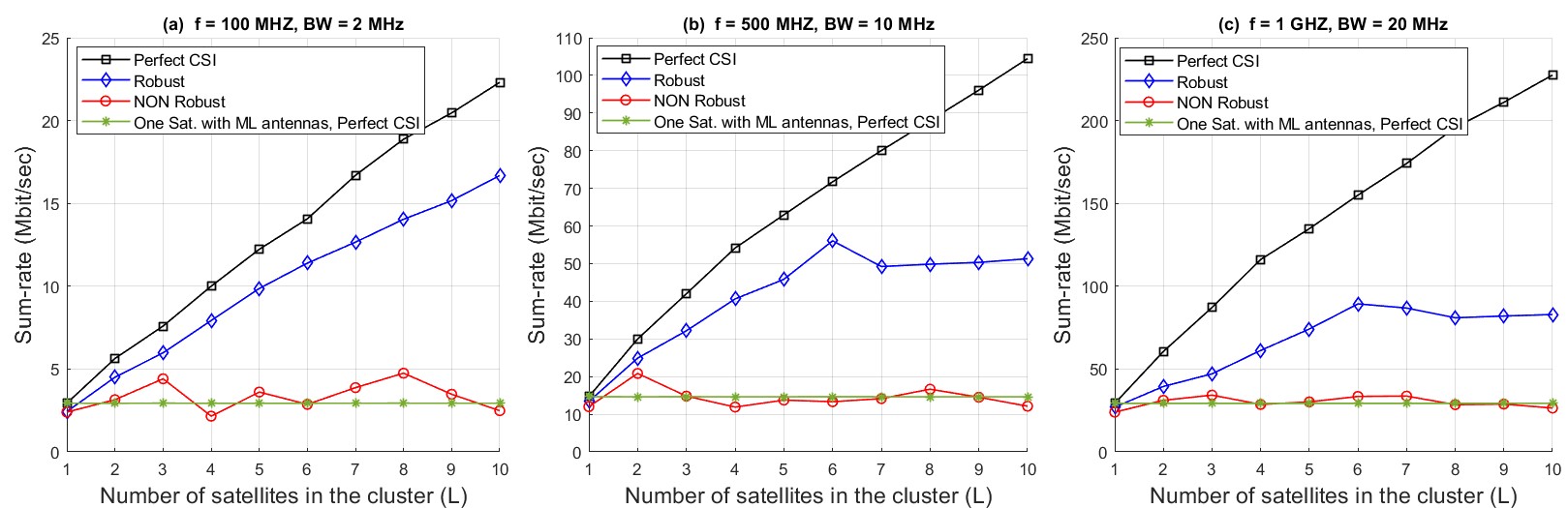}
    \caption{Sum-rate vs $L$, in different frequencies and bandwidths.}
    \label{fig:all frequencies}
\end{figure*}
\section{Numerical Results}\label{Section Numerical Results}
In this section, we evaluate the performance of the proposed system through simulations. 
We assume that enough satellites with high processing power are launched and are in orbit and hence at each time, a NC satellite is visible to users on the ground. 

For our simulation, we consider a constellation of satellites each containing a UPA of $M=3\times 3$ elements dedicated to this specific regional transmission, with the transmit power limit of $P_l=10$, $l\in\{1,...,L\}$. We considered a constellation similar to Starlink, with four active layers and a total of $4236$ satellites in orbit. The constellation parameters are given in Table \ref{Table: Constellation Parameters}. The number of single-antenna users covered by the cluster of satellites is $K=10$, and they are randomly located in a $50$ km radius area in the Lake District National Park, UK (with latitude of $54.5260000^{\degree}$ and longitude of $-3.3000000^{\degree}$). The maximum number of visible satellites from these coordinates is $L_{max}=43$ satellites which are within the $30^{\degree}$ elevation angle. However, in this study we only use the nearest $L=10$ satellites to form a cluster. We select the nearest satellites because their channels undergo the least amount of path-loss and delay. Assuming that the delay is around $T_d=\frac{d_{max}}{c}=1.9$ milli-seconds, and the pilot transmission rate is $0.1$ Mbit/sec, the {CSI}  (${\mb{G}}$) is generated every $10$ micro-seconds, based on the model presented in (\ref{Channel Model}). At each time step $n$, the estimated channel $\hat{\mb{G}}(n)$ is set to the channel at the time step $\mb{G}(n-n_{\text{delay}})$, where  {$n_{\text{delay}}=\frac{1.9 \text{ milli-seconds}}{10\text{ micro-seconds}}=190$} in our setup. Note that the value of $n_{\text{delay}}$ is the same for all frequencies. 
{The channel uncertainty is defined as $\tilde{\mb{G}}(n) = \mb{G}(n-n_{\text{delay}}) -  \mb{G}(n)$ and the long term properties of channel uncertainty e.g. $\mathbb{E}\{\tilde{\mb{G}}_l\}$ and $\mathbb{E}\{\tilde{\mb{G}}_l\tilde{\mb{G}}_l^H\}$, $\forall l\in\{1,...,L\}$, are calculated based on this definition.}
% The long term properties of channel uncertainty which are needed for precoding, e.g. $\mathbb{E}\{\tilde{\mb{G}}_l\}$ and $\mathbb{E}\{\tilde{\mb{G}}_l\tilde{\mb{G}}_l^H\}$, $\forall l\in\{1,...,L\}$, are calculated in a window of  $n_W=30$ steps.
Finally, the receiver noise power is obtained by $\sigma^2={K_{B}\times T \times \text{BW}}$, where $K_{B}$ demonstrates the Boltzmann constant, $T=280$ is temperature in Kelvin, and $\text{BW}=0.02f$ is bandwidth in Hz.   

\begin{table}[t]
    \centering
    \begin{tabular}{|c|c|c|c|c|}
    \hline
         \textbf{Parameter} & \textbf{Layer1} & \textbf{Layer2} & \textbf{Layer3} & \textbf{Layer4} \\
         \hline
         Num. orbital planes & $72$ & $36$ & $6$ & $72$ \\
         \hline
         Num. satellites per plane & $22$& $20$& $58$& $22$ \\
         \hline
         Altitude &$550$ km&$570$ km&$560$ km&$540$ km  \\
         \hline
         Inclination angle & $53^{\degree}$ &$70^{\degree}$&$97.6^{\degree}$&$53.2^{\degree}$ \\
         
    \hline
    \end{tabular}
    \caption{Constellation Parameters}
    \label{Table: Constellation Parameters}
\end{table}

Fig. \ref{fig:all frequencies} demonstrates our simulation results, where we have shown the achievable sum-rate versus the number of satellites in the cluster. The performance of the robust technique is compared to the performance of the perfect CSI and non-robust scenarios. In the perfect CSI scenario it is assumed that $\mb{\hat{G}}(t)=\mb{G}(t)$, while in the non-robust scenario it is assumed that $\mb{\hat{G}}(t)=\mb{G}(t-T_d)$ but the NC is not aware of the long-term properties of CSI uncertainty and hence it assumes $\mathbb{E}\{\tilde{\mb{G}}_l\}=\mb{0}$ and $\mathbb{E}\{\tilde{\mb{G}}_l\tilde{\mb{G}}_l^H\}=\mb{0}$, $l\in\{0,...,L\}$. The simulations are provided in three different setups with different operating frequencies of $f\in\{100 \text{ MHz}, 500\text{ MHz}, 1\text{ GHz}\}$ and bandwidths of $BW=0.02f$. 
As expected, by increasing the frequency, the value of the CSI uncertainty rises and the algorithm shows less robustness to the outdated CSI. Thus, this technique is useful in frequency bands of up to $1$ GHz, which allows for the integration of some of the terrestrial communication systems with SatComs. 

We can see that by allowing more satellites to cooperate with each other, the performance of the system becomes better. Note that the satellites are added to the cluster based on their distances to the users, i.e. the first satellite has the shortest distance to the centre of the covered region on earth, the second satellite has the second shortest distance, and so on. Therefore, the first $10$ satellites are the ones with lowest path-loss values. If we add more satellites which are further from the users, the performance does not increase linearly, but the overhead of information sharing significantly increases in the cluster. Also, adding distant satellites to the cluster makes the scheduling harder, since the propagation delays would be much different; in other words, the nearest satellite should stay idle until the furthest satellite receives the pilot and can estimate the channels. This increases the CSI delay ($T_d$) to equal the delay of the furthest satellite in the cluster.

Finally, we compare the multi-satellite communication system with one satellite having the same number of antenna elements as all satellites in the cluster combined. We see that in our simulation setup, addition of more antennas on the same satellite does not change the performance. This is because of the high correlation between the channels of antennas at the same satellite which does not allow for diversity. If the channels of different antennas in one satellite experienced independent fading, the rate would have increased with the number of antennas on a single satellite. However, this is not the case considering the realistic channel model presented in (\ref{Channel Model}). Therefore, it can be concluded that the proposed robust distributed MIMO satellite system considerabley outperforms the single-satellite scenario with perfect CSI. 

%Changing the channel model to an i.i.d distribution would increase the rate with number of antennas on a single satellite, but that is not realistic. 
%Therefore, in a realistic scenario, with the channel model presented in (\ref{Channel Model}), we can see that the proposed robust distributed MIMO satellite system outperforms the single-satellite scenario with perfect CSI.  }

\section{Conclusion} \label{Section Conclution}
In this paper we investigate the integration of LEO SatCom with terrestrial communication. To be specific, we study the performance of multi-satellite MIMO communication system in presence of delayed CSI, and we propose a robust precoding technique to cope with the effects of channel uncertainty. Our proposed solution has very low computational complexity and it works well for frequency bands of up to $1$ GHz. Furthermore, it is shown that the presented multi-satellite technique outperforms single-satellite scenario with the same number of antennas as the entire cluster of satellites.

% \appendices
% \section{First }\label{FirstAppendix}

\bibliographystyle{IEEEtran}
\bibliography{MyRef}
\end{document}